\documentstyle[fleqn,twoside,epsfig,amssymb,STAS2001a]{article}

\newcommand{\be}{\begin{eqnarray}}
\newcommand{\ee}{\end{eqnarray}}
\newcommand{\bea}{\begin{eqnarray}}
\newcommand{\eea}{\end{eqnarray}}

\begin{document}

\setcounter{page}{225}

\label{flin1}

\heads{Katarzyna Bajan, Monika Biernacka, Piotr  Flin,
    W{\l}odzimierz God{\l}owski, Victor Pervushin, Andrey Zorin}
    {Large Scale Periodicity in Redshift Distribution}

\twocolumn[

\Arthead{flin1}{flin2}

\Title{LARGE SCALE PERIODICITY IN REDSHIFT DISTRIBUTION}

\cont{Katarzyna Bajan, Monika Biernacka, Piotr  Flin,
    W{\l}odzimierz God{\l}owski, Victor Pervushin, Andrey Zorin}
    {LARGE SCALE PERIODICITY IN REDSHIFT DISTRIBUTION}

\noi{\large\bf Katarzyna Bajan$\blacktriangle$, Monika Biernacka$\blacklozenge$,
    Piotr  Flin\foom 1{\small $\blacksquare$}, W{\l}odzimierz God{\l}owski$\bigstar$,\y\\
    Victor Pervushin$\clubsuit$, Andrey Zorin$\spadesuit$}\y\\

\noi\hg{-0.1}$\blacktriangle${\it H. Niewodniczanski Institute of
Nuclear Physics, 31-342 Krakow, ul.
        Radzikowskiego 152,  Poland}\y\\
$\blacklozenge${\it Pedagogical University, Institute of Physics, 25-406
        Kielce, ul. Swietokrzyska 15, Poland}\y\\
{\small $\blacksquare$}{\it Pedagogical University,
        Institute of Physics, 25-406 Kielce, ul. Swietokrzyska 15, Poland;\\
        \hg{0.8}Bogoliubov Laboratory of
        Theoretical Physics, Joint Institute for Nuclear Research,
        141980, Dubna, Russia}\y\\
\hg{-0.18}$\bigstar${\it Astronomical Observatory of the Jagiellonian University, 30-244
        Krakow, ul. Orla 171,Poland}\y\\
$\clubsuit${\it Bogoliubov Laboratory of Theoretical Physics, Joint Institute for Nuclear
        Research, 141980, Dubna, Russia}\y\\
$\spadesuit${\it Faculty of Physics, MSU, Vorobjovy Gory, Moscow,
        119899, Russia}\\

\Rec{October 23, 2003} \Abstract{ We review the previous studies
of galaxies and quasar redshifts discretisation. We present also
the investigations of the large scale periodicity, detected by
pencil--beam observations, which revealed 128 (1/h) Mpc period,
afterwards confirmed with supercluster studies.  We present the
theoretical possibility of obtaining such a periodicity using a
toy-model. We solved the Kepler problem, i.e. the equation of
motion of a particle with null energy moving in the uniform,
expanding Universe, decribed by FLRW metrics. It is possible to
obtain theoretically the separation between large scale structures
similar to the observed one. }

]

\email 3 {sfflin@cyf-kr.edu.pl}



\section{Introduction}

In the large scale Universe the search of regularities is
connected with testing, if radial velocities of galaxies can admit
an arbitrary values, or some regular patterns, called
periodisation or quantisation of galaxy redshifts is observed. The
search for possible periodicity in redshift distribution has been
an important question from the point of view of selection effects
in observational data, as well as the correctness of applied
statistics. The possible interpretation of such effect is very
important too. The lack of a known mechanism causing redshift
discretisation constituted the basis for claims invoking new
physics at work. Due to different methods of investigations we
describe separately investigations of galaxy and quasar redshift
periodisation. There are presented in sections two and three. The
quasar redshift periodisation in more details is presented
elsewhere (Bajan et al. 2004) Moreover, the fourth section is
devoted to the large scale periodicity with a period of a 128
(1/h) Mpc. This large scale periodicity obtained originally by
pencil beam observations has been confirmed by various studies.
So, in reality, it is the only redshift discretisation, which is
almost commonly accepted. In the last section we give the possible
solution of the large scale periodicity. We obtain the exact
solution of the Kepler problem in the expanding Universe. With
appropriate fine tunning it is possible to obtain the distribution
of structures with about 130 Mpc separation between them.

\section{The discretisation of galaxy redshifts}

The discussion of redshifts quantisation started with works of Tifft.
In the Coma cluster of galaxies (Tifft 1976) optically determined
redshifts of member objects can be grouped into bands with the
differences of 72.46 km$\cdot$s$^{-1}$. The repetition of the analysis
based on the more accurate, this is HI, data confirmed this
periodicity.

The result was preceded by the study of the distribution of galaxy
redshifts versus nuclear magnitudes in the Coma cluster core
(Tifft 1974 and references therein), were strong redshifts
periodicity of 220 km$\cdot$s$^{-1}$ were observed. The existence
of band structure on nuclear magnitude-redshift diagram was
reported by Nanni et al. 1981 on the basis of better nuclear
magnitude and redshift determination.  However, in order to find
the effect of quantisation it was necessary to apply to galaxy
heliocentric redshift also the correction connected with the solar
galactocentric movement, which is not precisely known.

Few years later  Tifft and Cocke (1984) generalised their claim for all
galaxies. There exists global periodicity. It is equal to $36.3$ km$\cdot$s$^{-1}$
for broad-line galaxies and 24.2 km$\cdot$s$^{-1}$ for narrow-line galaxies.
The later value afterwards was not confirmed, so presently the value of about
36 (or 72) km$\cdot$s$^{-1}$ is considered as the basic one.

        Tifft (1980) shown that the redshift differences for double galaxies
have a statistically significant grouping  near multiples of 72 km
$\cdot$ s$^{-1}$. Afterwards, the initial sample of galaxies has
been enlarged and objects with more accurate redshift
determination ($\sigma < 25$ km$\cdot$s$^{-1}$) considered (Tifft
1982a,b). The periodicity of 72 km $\cdot$ s$^{-1}$ was observed
only for such well determined redshifts. Greater uncertainties
blurred the observed periodicity. The way, in which statistical
analysis has been performed was critisied by Newman, Haynes \&
Terzian (1989). Cocke and Tifft (1991) re-addressed this problem,
concluding that the approach was  statistically correct. Schneider
and Salpeter (1991) used the sample of 107 isolated binary
galaxies with very reliable  21 cm redshifts. They do not find the
multiplicity of 144, 216 ... km $\cdot$ s$^{-1}$, suggested by
Tifft, but they confirm the excess in velocity differences near 72
km $\cdot$ s$^{-1}$.

  Considering 40 members of the Local Group Rudnicki et al (2001) claimed
the existance, at 95 \% significance level, of redshifts
periodisation  in the considered sample, however without the
precise value of the periodisation. They concluded, that the
uncertainties in the published redshift determination do not
allowed to make more precise  statements. It is worthwhile to
note, that the discretisation of redshift in the Local Group
concern not only galaxies, but also globular clusters. The further
analysis of the problem (Godlowski et al. 2003) shown that the
strict quantisation is excluded, but a weak evidence  of period P
about 36 (or 24) km $\cdot$ s$^{-1}$ is observed.

 Guthrie and Napier (1991) from the database extracted  89 non-Virgo spirals with  galactocentric
redshifts < 1000 km$\cdot$ s$^{-1}$ and quoted redshift accuracy
$\sigma$ $\leq$  4 km $\cdot$ s$^{-1}$.
An evidence of periodicity about 37.5 km $\cdot$ s$^{-1}$ was  noted.
Afterwards they (1996) repeated the analysis  using 247 spirals with galactocentric
radial velocities  < 2600 km$\cdot$ s$^{-1}$, finding strong periodisation with
P $\sim$ 37.6 km$\cdot$ s$^{-1}$.
The paper precisely describes the procedure  of Power Spectrum Analysis,  the hypothesis tested, as
well as  the determination of the solar vector in respect to  the Galaxy centre.

\section{The quasar redshift quantisation}

This phenomenon is probably the eldest one, because in the late
sixties of the previous century first claims were published. Cowan
(1968) reported an effect with a period of 0.1666 in redshift z =
$\Delta$$\lambda$/$\lambda$. The result was broadly discussed and
the re-analysis of existing data, with increasing number of
quasars considered,  was performed. The Power Spectrum Analysis
was implemented for periodicity search, which allows to determine
precisely  the statistical significance of the effects ( Lake \&
Roeder 1972).  Some further claims were that periodicity is 0.089
in log(1+z), which corresponds to  a factor 1.227 between values
of two consecutive maxima in the 1+z redshift distribution.
Kjaergaard (1978)  carefully discussed the manner, in which
quasars are found and concluded, that the strong selection effects
influence the observed redshift distribution, causing the
existence of several peaks. So, according to his conclusions
 observed peaks are due
 to selection effects.
Quite recently the sample of 1647 quasars selected
in the uniform way has been analysed. Some maxima can be observed, but
according to Power Spectrum Analysis they are statistically
insignificant (Hawkins et al. 2002).

\section{Large scale periodisation}

Broadhurst et al. (1990)  found the high peaks  separated by regular
spacing,
strongly indicating the existence of a 128 $\cdot$h$^{-1}$ Mpc
(where  h$^{-1}$
is the Hubble constant  in  units of 100 km $\cdot$ s$^{-1}$)
period in
the redshift distribution. This result based on narrow pencil-beam method
  was confirmed by other
researchers, who using the distribution of Abell galaxy clusters founded
a 120 Mpc
periodicity  (Einasto et al. 1997). Several deep galaxy redshift surveys
show this periodicity  the presence
 of 100-Mpc scale structures (Kirshner 1997). The possibility of such
 large scale distribution of matter was pointed out also by Gonzales et. al
 (2000).

The Voronoi foam being a model describing of galaxy distribution
in sheets, filaments and clusters surrounding voids was applied
for study the possibility of explanation this periodicity. Van de
Weygaert (1991) was able to explain the effect, but he concluded
that the periodicity, even in the cellular Universe, is not a
common phenomena. The similar conclusion was reached by Ikeuchi
and Turner (1991), who stressed that the quasi-periodicic pattern
is less than only  $2\sigma$ fluctuation in a
 Voronoi tessellation.

 The other simply possibility is the existence of a
coherent
peculiar velocity field  with the amplitude of the order of
10$^{-3}$ and the
period 128 h$^{-1}$ Mpc (Hill, Steinhardt and Turner 1991).
Such an amplitude is similar to
the peculiar velocity field observed in our vicinity and connected with
so-called
Great Attractor. The oscillating Universe (Morikawa 1990) or
the change
of the values of the fundamental constants (Hill, Steinhardt and Turner
(1990)) were
also involved as possible explanation of this periodicity.
Salgado et al. (1996), Quevedo et al. (1997) discussed the
oscillation of gravitational constant.

\section{The possible explanation of the large scale periodicity}

We study the Newtonian motion of a particle in a gravitational
field in the space--time with the
Friedmann---Lem\^aitre---Robertson---Walker (FLRW) metrics
\be\label{gr5} (ds^2)=(dt)^2-\sum_ia^2(t)(dx^i)^2. \ee The
observational coordinates $X^i$ of the expanding Universe can be
written as
 \be \label{fc} X^i=a(t)x^i,~~~dX^i=a(t)dx^i+x^ida(t),
 \ee
and instead of the differential of the Euclidean space $dX^i$,
we use the covariant differential of the FLRW space coordinates
\be\label{A0}
a(t)dx^i= d[a(t)x^i]-x^ida(t)=dX^i-X^i\frac{da(t)}{a(t)}.
\ee

One can check that the interval (\ref{gr5}) in terms of these variables
(\ref{fc}) becomes
 \be\label{A}
 (ds^2)=(dt)^2-\sum_i\left(dX^i-H(t)X^idt\right)^2,
 \ee
where $H(t)=\dot a(t)/a(t)$ is the Hubble parameter. As example
the classical Newton action in the space with the interval
(\ref{A}) and the covariant derivative $(\dot X^i-H(t)X^i)$  takes
the following form
\begin{equation}
\label{cr11a}
\hg{-1}S_A\hg{-0.3}=\hg{-0.5}\int\limits_{t_I}^{t_0}\hg{-0.5}dt\hg{-0.4}\left[\sum_i\left(P_i(\dot X^i-H(t)X^i)-
 \frac{P_i^2}{2m_I}\right)\hg{-0.3} +\hg{-0.3}\frac{\alpha}{\sqrt{X_iX^i}}\right]\hg{-0.5},
\end{equation}
 where $\alpha={M_{\rm O} m_I G}$ is a  constant of a Newtonian
 interaction of a galaxy with a mass $m_I$ in a gravitational field
  with central  mass ${M_{\rm O}}$.

In the terms of the conformal time $d\eta=dt/a$ in effective units
 $r=\sqrt{x_ix^i}=R/a$,
$P_r=P_Ra$ the
 action (\ref{cr11a}) takes the form \be
 S_A=\int\limits_{\eta_I}^{\eta_0}d\eta\left[P_r\frac{dr}{d\eta}
 +P_\theta\frac{d\theta}{d\eta}-\frac{P_r^2+{P_\theta^2}/{r^2}}{2m}
 -\frac{\alpha}{r}\right],\hskip-4pt
 \ee
where $m=a(\eta)m_I$.

Now let us consider a probably particle moving in a plane in the
cylindrical coordinates
 \be
 \label{coord}
 X^1=R\cos\Theta,~~X^2=R\sin\Theta
 \ee in the gravitational field  with Schwarzschild's
 metrics
\begin{equation}
\hg{-0.8}ds^2= \left(1-\frac{2\alpha}{m
r}\right) dt^2 -\frac{dr^2}{1-{2\alpha}/{(m r)}}-r^2\sin(\theta)^2d\theta^2.
\end{equation}

In this paper we consider the case of rigid state
when densities of a energy and a pressure are equals.
In this case $a(\eta)=\sqrt{1+2H_I(\eta-\eta_I)}$,
where the $H_I$ is the initial value of Hubble velocity (Behnke et. al 2002).  In the terms
of the conformal time $d\eta=dt/a$ in relative units $r=R/a$,
$P_r=P_Ra$ the action takes
the form:
\[
S_{\rm schw}=\int\limits_{\eta_I}^{\eta_0} d\eta
\left[ P_r\frac{dr}{d\eta} + P_\theta\frac{d\theta}{d\eta} - \right.
\]
\begin{equation}
\label{S:eta}
\hg{3}\left.-Q_{\rm
schw}\sqrt{P_r^2Q_{\rm schw}^2+P_\theta^2/r^2+m^2} + m\right],
\end{equation}
where $\displaystyle Q_{\rm
schw}=\left(1-\frac{r_g}{r}\frac{m_I}{m}\right)^{1/2} $, $r_g={M_{\rm O}  G}$,
and $P_r$, $P_\theta$ are conjugate impulses to corresponded coordinates. The
path of the probably particle is shown on the Fig.~1.

\begin{figure}\label{fig:bh}
    \centering
    \epsfig{figure=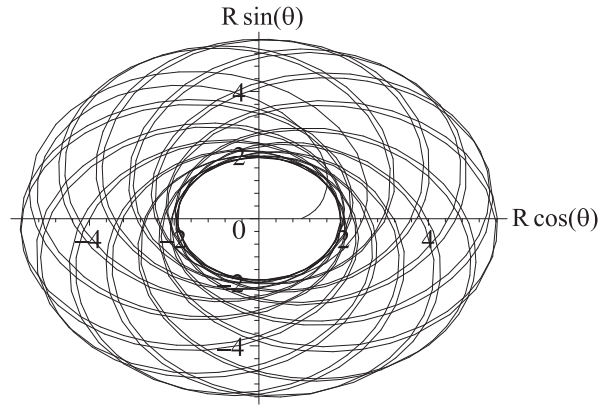,width=238pt}
    \caption{\protect\small
Solution equation of motion for the action (\ref{S:eta})
at
$r(\eta_I)=1$, $r'_{\eta}(\eta_I)=0$, and $\theta(\eta_I)=0$ for
$H_I=1$, $m_I=1$, $R_I=1$, $P_\theta=1$, and $r_g=0.5$. For these
values of initial data and integrals of motion the total system
energy (defined from (\ref{S:eta})) is equal to zero at initial
point $\eta=\eta_I$. This point corresponds to initial point of
the structure creation.
}
        \medskip\hrule
\end{figure}

We chose the total energy of a system defined in (\ref{S:eta}) as \be E=Q_{\rm
schw}\sqrt{P_r^2Q_{\rm schw}^2+P_\theta^2/r^2+m^2}- m \ee at the initial point
$\eta=\eta_I$ as equals zero. The trajectory begin at the point $(1, ~0)$. From
Fig. \ref{fig:bh} we see that some region are more occupied then others. In
such a way we show that in the expanding FLRW metrics and assuming rigid state
of matter, where density equals pressure, it is possible to obtained the
formation of clusters from the uniform distribution of particles with zero
gravitational distributional energy. It is possible to regard this mechanism as
a main reason of cellular structures is the Universe. With the fine tunning we
are able to find the distributions equals to about $130$ Mpc.

\section{ Conclusions}

There are several effects, which are called the redshifts
quantisations. The scale of this effect is drastically different.
From the almost local values to fantastically great part of the
Universe. The clear interpretation is only in the case of 100--Mpc
periodisation, which is probably  due to the large scale
structure, but such structure origin
 is not easily  explained. The discretisation of the quasar redshifts
 seems to be an artifact of selection effects and smallness of the
 samples. On the other hand Zhuck et al. (2001) show that the quasars
 seems to be located in thin walls of web with the size of $50 \div 100$ Mpc.
 This finding is in good agreement with the size of voids, as well as
 not in great contradiction with the large scale distribution of
 galaxies revealed by optical data.

   In the case of 36 (or 72) km$\cdot$s$^{-1}$
 periodisation, which is of
 the main interest of the   adherents of
 redshifts quantisation, we express
 the opinion that the effect is not
 convincingly shown.
 It is not quite clear, why from
  several thousands of galaxies
  in the Local Supercluster
only about 250 objects (i.e. less than 4\%) is taken into account,
which can caused a serious bias.  It will be interesting to
compare the real differences  between various galaxy redshifts
determination and not only to consider to quoted by authors formal
errors. This point was stressed also by Guthrie and Napier (1996).
If this alleged effect will be confirmed, this will open totally
new areas of investigations, opening also doors for so-called new
physics. Beforehand, however, the detailed analysis of
observational errors and applied statistical methods should be
performed. The papers presenting the results have obtained till
now clearly shown the weak and the strong points, which will be
fruitful in further work. Moreover, the investigations of the
redshift distribution in other superclusters will be of great
interest. Due to accumulation of precise measurements of redshifts
in the last years the re-discussion of the redshift discretisation
is possible and could help to solve this bizarre problem. We show
that the new physics is not needed for explanation of the
existence of the large scale periodicity of about 130 Mpc. Using
very simple model in which  a massive particle is circulating
around the source of gravitational field in the expanding Universe
with FLRW metrics it is possible to obtain regions with higher
density. This means that  particles will be with greater
probability located in some regions than others. With appropriate
fine tunning it is possible to obtain the separation between dense
regions (structures) of the order of 130 Mpc, which is consistent
with observations.

\label{flin2}

\clearpage

\end{document}